\documentstyle[prd,twocolumn,aps]{revtex}

\begin{document}
\draft
\preprint{   }
\title{RELATIVISTIC HYDRODYNAMICS IN CLOSE BINARY SYSTEMS:
ANALYSIS OF NEUTRON-STAR COLLAPSE}

\author{G. J. Mathews  and P. Marronetti}

\address{
University of Notre Dame,
Department of Physics,
Notre Dame, IN 46556}

\author{J. R. Wilson}
\address{
University of California,
Lawrence Livermore National Laboratory,
Livermore, CA  94550} 
\address{
University of Notre Dame,
Department of Physics,
Notre Dame, IN 46556}

\date{\today}
\maketitle
\begin{abstract}
We discuss the underlying relativistic physics which causes neutron
stars to compress and collapse in close binary systems as
has recently been observed in numerical (3+1) dimensional general
relativistic hydrodynamic simulations.
We show that compression is driven by velocity-dependent
relativistic hydrodynamic terms which increase the self gravity
of the stars.  They also  produce fluid motion
with respect to the corotating frame of the binary.
We present numerical and analytic results which confirm
that such terms are insignificant for uniform translation
or when the hydrodynamics is constrained
to rigid corotation.
 However, when  the hydrodynamics is 
unconstrained, the neutron star fluid relaxes
to a compressed  nonsynchronized state of almost no net intrinsic spin
with respect to a distant observer.  
  We also show that tidal decompression effects are 
much less than the velocity-dependent compression terms.
We discuss why several recent attempts to analyze this
effect with constrained hydrodynamics  or an analysis
of tidal forces do not observe compression.
We argue that an independent  test of this
must include unconstrained  relativistic
hydrodynamics to sufficiently high order 
that all relevant velocity-dependent terms and their
possible cancellations are included.

\end{abstract}
\pacs{PACS Numbers: 04.20.Jb, 04.30.+x, 47.75+f,, 95.30.Lz, 95.30.Sf
97.60.Jd, 97.80.-d}

\narrowtext

\section{INTRODUCTION}
\label{sec:level1}
The physical processes occurring during the last orbits of
a neutron-star binary are currently a subject of intense interest
\cite{wm95}-\cite{sbs98}.
In part, this recent surge in interest stems from 
relativistic numerical hydrodynamic simulations
in which it has been noted \cite{wm95,wmm96,mw97}  that as the
stars approach each other their interior density increases.
Indeed, 
for an appropriate equation of state, our numerical simulations indicate
that binary neutron stars collapse individually toward black holes
many seconds prior to merger.  This compression effect would
have a significant impact on the anticipated
gravity-wave signal from merging neutron stars.  It could
also provide an energy source for cosmological gamma-ray bursts
\cite{mw97}.

In view of the  unexpected nature 
of this neutron star compression effect and its possible repercussions, 
as well as the
extreme complexity of strong field general relativistic hydrodynamics,
it is of course imperative that there be an independent confirmation of the
existence of neutron star compression before one can  be convinced
of its operation in binary systems.  In view of this it is
of concern that the initial numerical results  reported in
\cite{wm95,wmm96,mw97} have been called into question.
A number of recent papers 
\cite{lai,rs96,wiseman,shibata,lombardi,lw96,brady,flanagan,thorne97,baumgarte,sbs98} 
have not observed this effect in Newtonian tidal forces \cite{lai}, 
first post-Newtonian (1PN) dynamics \cite{rs96,wiseman,shibata,lombardi,lw96,sbs98}, 
tidal expansions \cite{brady,flanagan,thorne97}, or in binaries
in which rigid corotation has been imposed \cite{baumgarte}.
The purpose of this paper is to point out that none of these recent studies
could or should have observed the compression effect which we
observe in our calculations.

Moreover, this flurry of activity has caused some confusion as 
to the physics to which we attribute the effects observed in the numerical
calculations.  The present paper, therefore,
summarizes our derivation of the physics which drives
the collapse.  We illustrate how such terms have been absent in some 
Newtonian or post-Newtonian
approximations to the dynamics of the  binary system.
We also present numerical results and analytic expressions
which demonstrate how
the compression forces result in an orbiting dynamical system
from the presence of fluid motion with respect to the 
corotating frame.  As such, they could not appear in an
an analysis of relativistic external tidal forces no matter
how many orders  are included in the tidal expansion parameter
(e.g.~\cite{flanagan,thorne97}) unless self gravity from internal
hydrodynamic motion is explicitly accounted for.
The effect could also not arise in systems with uniform translation
or rigid corotation.

The implication of the present study is that any attempt to
confirm or deny the compression
driving force requires an unconstrained, untruncated
relativistic hydrodynamic treatment.  At present,
ours is still the only existing such calculation.  Hence,
despite claims to the contrary 
\cite{lai,rs96,wiseman,shibata,lombardi,lw96,brady,flanagan,thorne97,baumgarte,sbs98},
 the neutron star compression effect
has not yet been independently tested.

Another confusing aspect surrounding the numerical results
has been our choice of a conformally flat
spatial three-metric for the solution of the field equations.  
Indeed, it
has been speculated that this approximate gauge choice (in
which the gravitational radiation is not explicitly manifest)
may have somehow led to spurious results.  A second purpose
of this paper, therefore, is to emphasize that the compression driving terms
are a completely general result from the
relativistic hydrodynamic
equations of motion.  The advantages of the conformally flat condition
are that the algebraic form of the compression driving
terms is easier to identify and that
the solutions to the field equations obtain a simple form.
It does not appear to be the case, however, that the imposition of
a conformally flat metric drives the compression.  It has
been nicely demonstrated in the work of  
Baumgarte et al.~\cite{baumgarte} that conformal flatness does not necessarily 
lead to neutron-star compression.

\section{The Spatially Conformally Flat Condition}
There has been some confusion in the literature as to
the uncertainties introduced by imposing a conformally flat condition
(henceforth {\it CFC}) on the spatial three-metric.  Therefore 
we summarize here some attempts which we
and others have made to quantify the nature of this approximation.

The only existing strong field numerical relativistic hydrodynamics
results in three unrestricted  spatial dimensions 
to date have been derived in the context of the
{\it CFC} as
described in detail in \cite{wm95,wmm96,mw97}.

We begin with the usual ADM (3+1) metric \cite{adm62,york79}
in which there is a slicing of the spacetime into a one-parameter
family of  three-dimensional hypersurfaces 
separated by differential displacements
in a timelike coordinate,
\begin{equation}
ds^2 = -(\alpha^2 - \beta_i\beta^i) dt^2 +
2 \beta_i dx^i dt + \gamma_{ij}dx^i dx^j~~,
\label{metric}
\end{equation}
where we take Latin indices to run over  spatial coordinates
and Greek indices to run over four coordinates. We also utilize 
geometrized units ($G = c = 1$) unless otherwise noted.
The scalar $\alpha$ is called the lapse function, $\beta_i$ 
is the shift vector,
and $\gamma_{i j}$ is the spatial three metric.

In what follows, we make use of the general relation between
the determinant of the four metric $g_{\alpha \beta}$ and the ADM metric
coefficients
\begin{equation}
det(g_{\alpha \beta} ) = - \alpha^2 det({\gamma_{i j}}) \equiv   \alpha^2 \gamma^2~~,
\end{equation}
 where $\gamma \equiv \sqrt{-det(\gamma_{i j})}$.

The conformally flat metric condition  simply expresses the 
three metric of Eq.~(\ref{metric}) as a position
dependent conformal factor $\phi^4$ times a flat-space Kronecker delta 
\begin{equation}
\gamma_{i j} =  \phi^4 \delta_{ij}~~.
\end{equation}

It is common practice (e.g.~\cite{evans85,cook93,brugmann97})
to impose this condition when solving the initial value problem
in numerical relativity.  It is the natural choice
for our  three-dimensional
quasiequilibrium orbit calculations \cite{wmm96} which in essence
seek to identify a sequence of initial data configurations
for neutron-star binaries.

The reason conformal flatness is chosen most frequently
for the initial value problem is that it simplifies the solution of
the hydrodynamics and field equations.
The six independent components
of the three metric are reduced to a single
position dependent conformal factor.

Since conformal flatness  implies no 
transverse traceless part of $\gamma_{i j}$ it can
minimize the amount of initial gravitational radiation 
apparent in the initial configuration.
However, in general  the physical data
still contain a small amount of
preexisting gravitational radiation.
This has been clearly demonstrated in numerical calculations of
axisymmetric black-hole collisions \cite{smarr}.  
In exact numerical simulations,
 the gravitational radiation appears as the
time derivatives of the spatial three metric 
($\dot \gamma_{i j}$)  and its conjugate (the extrinsic curvature 
$\dot K_{i j}$) are evolved.  
The immediate evolution of the fields
from conformally flat initial data
is characterized by the development of a weak 
gravity wave exiting the system.

An estimate
of the radiation content of initial data slices  for
axisymmetric black hole collisions has been made
by Abrahams \cite{abrahams}.  Even for high values of momentum,
the initial slice radiation is always less than about 10\% 
of the maximum possible radiation 
energy (as estimated from the area theorem).  

Two questions then are relevant to our application of
the {\it CFC}.  One is the validity
of this metric choice for the initial value  problem, 
and the other
is the effect on the system of the "hidden" gravitational
radiation in the physical data.

Regarding the validity of the {\it CFC}
one has a great deal of freedom in choosing
coordinates and initial conditions as long as the 
initial space is Riemannian and the
metric coefficients satisfy the constraint
equations of general relativity \cite{mtw}.
Indeed, we have shown in \cite{wmm96} that 
exact solutions for the {\it CFC} metric
coefficients can be obtained by imposing the ADM
Hamiltonian and momentum constraint conditions.  Nevertheless,
in three dimensions a physical space is conformally flat if
and only if the Cotton-York tensor vanishes \cite{kramer80,eisenhart},
\begin{equation}
C^{i j} = 2 \epsilon^{i k l}\biggl( R^j_{~k}
- {1 \over 4} \delta^j_{~k}R \biggr)_{;l}~~,
\label{cy}
\end{equation}
where $R^j_{~k}$ is the Ricci tensor and $R$ is the Ricci scalar
for the three space.

Equation (\ref{cy})  vanishes by fiat for the three-space metric we have
chosen. However, conformally flat
solutions for physical problems have only been proven \cite{kramer80,eisenhart}
 for spaces of special symmetry
(e.g.~constant curvature, spherical symmetry, time symmetry, 
Robertson-Walker, etc.  \cite{kramer80}).
Hence, the invocation of the CFC here and in other applications
is an assumption.  That is, it is a valid solution to
the Einstein constraint equations, but does not necessarily
describe a physical configuration to which two neutron stars
will evolve.   Nevertheless, this is a valid approximation as long
as the nonconformal contributions from the
$\dot \gamma_{i j}$ and $\dot K_{i j}$ equations in the exact two-neutron
star problem remain small.
Indeed, numerical tests for an axisymmetric
rotating neutron star \cite{cookcfa} and 
a comparison of the {\it CFC} vs.~an exact metric expansion
for an equal-mass binary
 \cite{rs96} have indicated that conformal flatness is 
a good approximation when it can be tested.

As a related illustration, consider the Kerr 
solution for a rotating black hole.  It is well known that the Kerr
metric is not conformally flat.  The close binaries
we study have specific angular momentum only slightly
greater than that of an extreme Kerr black hole.
 Also, they ultimately merge and collapse to a single  Kerr black hole.
Hence, an analysis of the Cotton-York tensor for a Kerr black hole is
another indicator of the degree to which conformal flatness
is a valid approximation for neutron-star binaries. 

Figure \ref{fig1}.  
gives the dimensionless scaled Cotton-York parameter
$C^{\theta \phi} m^3$ for 
a maximally rotating Kerr black hole as a function of
proper distance.  For illustration, consider the decrease of this
quantity as one moves away from the horizon at $m = r$ as a measure
of the rate at which the metric becomes conformally flat.
The maximally rotating
($a = m$)  black hole of this example, however,  
is an extreme example
of compactness and angular velocity relative to any of the neutron
stars in our simulations.  

It can be seen in figure \ref{fig1} that, even for this extreme case,
the dimensionless 
tensor coefficient
$C^{\theta \phi} m^3$ diminishes rapidly away from the black hole.
At the separation of interest for binary neutron stars approaching 
their final orbits
($r/m \sim 25$ where $m$ is the total binary mass and $r$ the separation
between stars), this coefficient has already diminished to
$\sim 10^{-3}$ of the value at the event horizon, ($r/m \sim 1$). 
Thus, the effect of either star on its companion
is probably well approximated by conformal flatness. 
Regarding the interior of the neutron stars themselves, in our studies
the stars are rotating so slowly (even when corotating) that the
deviation from conformal flatness is probably negligible.
Thus, it seems plausible that
conformal flatness is  a 
reasonable approximation for most physical aspects
involving the spatial three-metric of binary neutron-star systems.

The next issue concerns the "hidden" radiation in the physical
data.  To address this we decompose the extrinsic curvature
into longitudinal $K^{i j}_L$ and 
transverse $K^{i j}_T$ components as proposed by York \cite{york73},
\begin{equation} 
K^{ij} = K^{i j}_L + K^{i j}_T~~.
\label{kdecomp}
\end{equation}
By definition the transverse part obeys
\begin{equation}
D_i  K^{i j}_T  = 0~~,
\label{kt}
\end{equation}
where $D_i$ are covariant derivatives.  
The longitudinal part can be derived from a properly
symmetrized vector potential.  We find 
\begin{equation}
D_i  K^{i j}_L  = 8 \pi S^i~~,
\label{kl}
\end{equation}
where $S^i$ are spatial components of the contravariant
four-momentum density.

The product $K^{i j}_T K_{Tij}$ is a measure of the 
hidden radiation energy density.
To find $K^{i j}_T$ then from our numerical calculations,
 we first find  $K_{i j}$ by choosing maximal slicing 
[$Tr(K_{i j}) = 0$] and requiring that the trace free part
of the  $\dot \gamma_{i j}$ equation vanish.  This gives \cite{wmm96}  
\begin{equation}
2\alpha K_{ij} = (D_i \beta_j+D_j \beta_i -{2\over 3} \phi^{-4} 
\delta_{ij} D_k \beta^k)~~.
\label{detweiler}
\end{equation}
We then determine $K^{i j}_L$ from the equilibrium momentum density 
[Eq.~(\ref{kl})] and  subtract $K^{i j}_L$ from $K^{i j}$.

We find that this measure of the "hidden" gravitational radiation
energy density is a small fraction of the total gravitational
mass energy of the system,
\begin{equation}
\int  K^{i j}_T K_{Tij} {dV\over 8 \pi} \approx 2 \times 10^{-5} ~
{\rm M_G}~~.
\end{equation}
Hence, we conclude that the CFC is probably a good
approximation to the initial data.  

This should be an excellent approximation for the determination
of stellar structure and
stability. However, 
an unknown uncertainty  enters if one attempts to 
reconstruct the time evolution
of the system (e.g.~the gravitational waveform)
from this sequence of quasistatic initial conditions.
At present we make this connection 
approximately via a multipole expansion \cite{thorne}
for the gravitational radiation as described in  \cite{wmm96}.

\subsection{An Electromagnetic Analogy}

The meaning of imposing a conformally flat spatial metric
can, perhaps, be qualitatively understood in an electromagnetic analogy.
Both the ADM formulation of relativity and Maxwell's equations
can be written as two constraint equations plus two dynamical
equations.  In electromagnetism the constraint equations 
for electric and magnetic fields
are embodied in the $\nabla \cdot E$ and $\nabla \cdot B$ equations,
while the dynamical equations are contained in Ampere's 
law and Faraday`s law.
In relativity the analogous  constraint equations are the 
ADM momentum and Hamiltonian
constraints. The dynamical equations are 
the ADM $\dot K_{i j}$ and $\dot \gamma_{i j}$
equations.  In either electromagnetism or gravity,
any field configuration which satisfies
the constraint equations alone represents a 
valid initial value solution.  However, one must
analyze its physical meaning.

  For example, consider two orbiting charges.
One could construct an electric field which satisfies the constraint
by simply summing over the electrostatic field
from two point charges. Similarly,
one can construct a static magnetic field from  the charge current
by imposing $\dot E = \dot B = 0$ in the dynamical equations.
However, by forcing the dynamical equations to vanish, one has
precluded the existence of electromagnetic radiation.
In this field configuration, therefore one has unknowingly imposed 
ingoing radiation to cancel the 
outgoing electromagnetic waves.  

Similarly,  enforcing 
$\dot K_{i j}$ = $\dot \gamma_{i j} = 0$
might in part be thought of as implying the existence of
ingoing gravitational
radiation to cancel the outgoing gravity waves.  Nevertheless, in
both cases, this remains a good approximation to the
physical system (with no ingoing wave) as long as the
energy density contained in the radiation is small compared
to the energy in orbital motion.  

Gravity waves enter in two ways: as estimated above there is 
an insignificant  amount of "hidden"  radiation induced by our choice of 
the CFC;   there is also  the emission of
gravitational radiation by the orbiting binary system. The binary
gravity-wave emission is estimated in our calculations 
by evaluating the multipole moments and
using the appropriate formulas \cite{wmm96}. The fractional 
energy and angular momentum
loss rate as determined by the multipole expansion method is quite
small, e.g.~$\dot J/\omega J \sim 10^{-4}$
in all of our calculations \cite{wmm96,mw97}.
Hence, it can be concluded that the energy in gravitational
radiation is indeed small compared
to the energy in orbital motion.

  The emission of gravity
    waves also induces a reaction force which we have 
incorporated into our
hydrodynamic equations by the quadrupole formula. 
The radiation reaction force is so small, however, that
it is difficult to discern it in the numerical results.   In most of our
calculations we simply neglect the back reaction terms and thereby 
obtain quasistatic orbit solutions.

\subsection{Solutions to Field Equations}

With a conformally flat metric, 
the constraint equations for the field variables
$\phi$, $\alpha$, and $\beta^i$  reduce to simple Poisson-like equations
in flat space.
The Hamiltonian constraint equation \cite{york79}
for the conformal factor $\phi$ becomes \cite{wmm96,evans85}, 
\begin{equation}
\nabla^2{\phi} = -2\pi
\phi^5 \biggl[ (1+U^2) \sigma  - P 
+ {1 \over 16\pi} K_{ij}K^{ij}\biggr]~.
\label{phieq}
\end{equation}
 where $\sigma$ is the inertial mass-energy density
\begin{equation}
\sigma \equiv \rho (1 + \epsilon) + P  ~~,
\end{equation}
and $\rho$ is the local proper baryon density which is simply
related to the baryon number density $n$, $\rho = \mu m_\mu
n/N_A$, where $\mu$ is the mean molecular weight,
$m_\mu$ the atomic mass unit, and $N_A$ is Avogadro's number.
{$\epsilon$} denotes the internal energy per
unit mass of the fluid, and $P$ is the pressure. 
In analogy with special relativity we have also introduced
a Lorentz-like variable
\begin{eqnarray}
\biggl[ 1 + U^2\biggr]^{1/2} \equiv  \alpha U^t  &= &\biggl[ 1 + U^j U_j \biggr]^{1/2} 
\nonumber \\
&&= \biggl[ 1 + \gamma^{i j}  U_i U_j \biggr]^{1/2}~~,
\label{weq}
\end{eqnarray}
where $U_i$ is the spatial part of the covariant four velocity.
Here we explicitly write $U^2$ 
(in place of $W^2 - 1 $ used in 
\cite{wm95,wmm96,mw97}) because
it emphasizes the extra velocity dependence  here and in the 
equations of motion.

In the Newtonian limit, the r.h.s.~of Eq.~(\ref{phieq}) 
is dominated \cite{wmm96} by the
proper matter density $\rho$, but in relativistic neutron stars
 there are also
contributions from the internal energy density $\epsilon$,
pressure $P$, and extrinsic curvature.  This Poisson  source is
also enhanced by the  generalized
curved-space Lorentz factor $(1+U^2)$.
 This velocity  factor becomes important as the 
orbit decays deeper into the gravitational potential and the
orbital kinetic energy of the binary increases.

It was pointed out in  the appendix of \cite{mw97} 
that in analogy to the velocity-dependent enhancement of
the source for Eq.~(\ref{phieq}),  the 
Poisson  source for the 
$v^4$ post-Newtonian correction to the effective potential
also exhibits velocity-dependence.  This
appendix has been misinterpreted as a statement that we
attribute the compression to a first post-Newtonian effect.
We therefore wish to state clearly that the appendix in that
paper was merely an illustration of how  the effective
gravity begins to show velocity dependence even in a post-Newtonian 
expansion.  
The velocity dependence of the post-Newtonian source is not
the main compression driving force.  The compression derives mostly
from the hydrodynamic terms described herein.
It is not obvious, however, at what post-Newtonian order
the compression effect should be counted, since different authors
have treated these terms differently.  We return to this point below.

In a similar manner \cite{wmm96}, the Hamiltonian constraint, 
together with the maximal slicing condition,
provides an equation for the lapse function,
\begin{eqnarray}
\nabla^2(\alpha\phi) = && 2 \pi
\alpha \phi^5 \biggl[ (3(U^2+1) \sigma  \nonumber \\
&& - 2 \rho(1 + \epsilon) + 3 P
 + {7  \over 16\pi} K_{ij}K^{ij}\biggr]~~.
\label{alphaeq}
\end{eqnarray}
Here again, the source is strengthened when the fluid is
in motion through 
the presence of a $U^2+1$ factor and the $K_{ij}K^{ij}$ term.

The momentum constraints \cite{york79} provide 
an elliptic equation \cite{wmm96} for the shift vector,
\begin{equation}
\nabla^2 \beta^i = {\partial \over \partial x^i} \biggl({1 \over 3}
\nabla \cdot \beta\biggr) + 4 \pi \rho_3^i~,
\label{wilson3}
\end{equation}
\begin{eqnarray}
\rho_3^i & =& \biggl(4\alpha \phi^4 S_i - 4 \beta^i(U^2+1)\sigma \biggr)
\nonumber \\
 && {1 \over 4\pi} {\partial  ln(\alpha/\phi^6) \over \partial x^j}
({\partial \over \partial x^j}\beta^i + {\partial \over \partial x^i} \beta^j
-{2\over 3}\delta_{ij} {\partial \over \partial x^k} \beta^k)~, 
\label{betaeq}
\end{eqnarray}
where  we have introduced \cite{w79} the 
Lorentz contracted
coordinate covariant momentum density,
\begin{equation}
S_i = \sigma W U_i~~.
\label{momdef}
\end{equation}

As noted previously and in Ref.~\cite{wmm96}, we only solve
equation  (\ref{wilson3})  for the
small residual frame drag after the dominant $\vec \omega \times \vec r$ contribution
to $\vec \beta$ has been subtracted.

\section{Relativistic Hydrodynamics}

The techniques of general relativistic hydrodynamics have
been in place and well studied for over 25 years \cite{w79}.
The basic physical
processes which induce compression can be traced to completely general terms
in the hydrodynamic equations of motion.
To illustrate this we first
summarize the completely general derivation of the 
relativistic covariant momentum equation in  Eulerian form and identify
the terms which we believe to be the dominant contributors to
the  relativistic compression effect.

For hydrodynamic simulations it is convenient to explicitly
consider two different
 spatial velocity fields.  One is $U_i$, the spatial
components of the covariant four
velocity. The other is $V^i$, the contravariant coordinate matter
three velocity, which is related to the four velocity 
\begin{equation}
V^i = { U^i \over U^t}  =
{\gamma^{i j} U_j \over U^t} - \beta^i~~.
\label{threevel}
\end{equation}
It is convenient to select the
shift vector $\beta^i$ such that the 
coordinate three velocity vanishes when averaged over the star,
$\langle V^i\rangle = 0$.
This minimizes coordinate fluid motion with respect to
the shifting ADM grid.

The perfect fluid energy-momentum tensor is
\begin{equation}
T_{\mu \sigma} = \sigma U_\mu U_\sigma
 + P g_{\mu \sigma}~~.
\end{equation}
However, it is convenient to 
derive the hydrodynamic equations of motion using
the mixed form,
\begin{equation}
T_\mu^{~\nu} =
g^{\sigma \nu} T_{\mu \sigma} = \sigma U_\mu U^\nu + P \delta_{\mu}^{~\nu}~~,
\end{equation}

The vanishing of the
spatial components of the divergence of the
energy momentum tensor
\begin{equation}
\biggr(T_i^{~\mu}\biggl)_{;\mu} = 0
\end{equation}
leads to an evolution equation for the spatial
components of the covariant four momentum,
\begin{eqnarray}
{1\over\alpha \gamma}{\partial (S_i \gamma )\over\partial t} +&&
{1\over\alpha \gamma}{\partial (S_i V^j\gamma )\over\partial x^j}
+{\partial P\over \partial x^i} \nonumber \\
&& +  {1 \over 2} {\partial g^{\alpha \beta}\over\partial
x^i} { S_\alpha S_\beta \over S^t} = 0
~~. 
\label{divmom}
\end{eqnarray}
The covariant momentum equation is used because of its
close similarity with Newtonian hydrodynamics.
The first two terms are advection terms familiar from
Newtonian fluid mechanics.  The latter two terms are the
pressure and gravitational forces, respectively.

Expanding the gravitational acceleration into
individual terms we have
\begin{eqnarray}
{\dot S_i}& + &  S_i{\dot \gamma \over\gamma}
+{1\over\gamma}{\partial\over\partial x^j}(S_iV^j\gamma)
 + {\alpha \partial P\over \partial x^i}
- S_j {\partial \beta^j \over \partial x^i}
\nonumber \\
 & + & \sigma {\partial \alpha \over \partial x^i}
 + \sigma  \alpha \biggl( U^2   {\partial \ln{\alpha} \over \partial x^i}
+ {U_j U_k \over 2 } {\partial \gamma^{j k}
\over \partial x^i}\biggr) = 0~~.
\label{hydromom}
\end{eqnarray}

Similar forms can be derived for the condition of baryon conservation
and the evolution of internal energy \cite{wmm96,w79}.  
However, the above momentum
equation is sufficient for the present discussion.

It is now worthwhile to consider the ''gravitational''
forces embedded in the expanded terms of 
Equation (\ref{hydromom}).  These 
result from the affine connection terms
$\Gamma^\mu_{\mu \lambda} T^{\mu \lambda}$ in the covariant differentiation
of $ T^{\mu \nu}$.

The term containing
$\partial \alpha/\partial x^i$ comes from the time-time
component of the covariant derivative.  It
is of course the well known analog of the
Newtonian gravitational force as can easily be seen in the
Newtonian limit $\alpha \rightarrow 1 - Gm/r$.  

The term $S_j (\partial \beta^j /\partial x^i)$
comes from the space-time covariant derivative.
In an orbiting system 
it is convenient to allow $\beta^j$ to follow the
orbital motion of the stars.  In our specific application
\cite{wmm96} we let $\vec \beta = \vec \omega \times
\vec R + \vec \beta_{resid}^{drag}$ where 
$\omega$ is chosen to minimize matter motion on the grid.
Hence, $\vec \omega \times
\vec R$ includes the major part of rotation plus
frame drag.  The quantity  $\vec \beta_{resid}^{drag}$
is the residual frame drag after subtraction of rotation
and is very small for the almost nonrotating stars which
result from our calculations.  
With $\beta^j$ dominated by $ \vec \omega \times
\vec R $, the term $S_j (\partial \beta^j /\partial x^i)$
is predominantly a centrifugal force.  

The $U^2  \partial \ln{\alpha} / \partial x^i$ term
arises from the time-time component 
of the affine connection piece of the covariant derivative.
The  $(U_j U_k /2 ) \partial \gamma^{j k}
/\partial x^i$ term similarly  arises from the space-space components.
They do not have a Newtonian analog.
As we shall see, these terms cancel when  a frame can
be chosen such that the whole fluid is at rest with respect
to the observer (or in the flat space limit).  However, 
for a star with fluid motion  in curved space, they
describe additional velocity-dependent  forces.

We identify the nonvanishing combination of these  
$U^2$-dependent  force terms 
and the  $S_j (\partial \beta^j /\partial x^i)$  
term as the major contributors to the net compression driving force.

This suggests some useful test problems for our hydrodynamic
simulations.  For example, in simple uniform translation
the effects of 
these terms must cancel to leave the
stellar structure unchanged.  Similarly, as
discussed below, any fluid motion
such that the four velocity can be taken as proportional
to a simple  Killing vector (e.g.~rigid corotation)
these force terms must cancel  \cite{baumgarte,kramer80}.
 However, for more general states of motion,  e.g.~noncorotating stars,
differential rotation, meridional
circulation, turbulent flow, etc., these  forces do not 
obviously cancel, but must
be evaluated numerically.  

Indeed, as discussed below,
the sign of these terms is such that
a lower energy configuration for the stars than that of rigid
corotation can be obtained by allowing the fluid to respond
to these forces.  As we shall see, the numerical
relaxation  of binary stars from corotation (or any other
initial spin configuration) produces 
a nonsynchronous (approximately irrotational) 
 state of almost no intrinsic neutron-star
spin in which the central density 
and gravitational binding energy increase.

\subsection{Conformally Flat Relativistic Hydrodynamics}
\label{hydro}

The practical implementation of conformal flatness means that,
given a distribution of mass and momentum on some manifold, we
first solve the
constraint equations of general relativity  at each time
for a given distribution of mass-energy.  
We then evolve the hydrodynamic equations to the next time step.
Thus, at each time slice we obtain
a solution to the relativistic field equations and
then can study the hydrodynamic response of the matter 
to these fields \cite{wmm96}. 

For the {\it CFC} metric, the relativistic momentum equation 
is derived by simply replacing $\gamma^{j k} \rightarrow \phi^{-4} 
\delta^{j k}$ in Eq.~(\ref{hydromom}).
\begin{eqnarray}
{\partial S_i\over\partial t}& +& 6 S_i{\partial \ln\phi\over\partial t}
+{1\over\phi^6}{\partial\over\partial x^j}\biggl(\phi^6S_iV^j\biggr)
+\alpha{\partial P\over \partial x^i} 
 -  S_j {\partial \beta^j \over \partial x^i} 
\nonumber \\
 & + & \sigma {\partial \alpha \over \partial x^i}
 + \sigma \alpha U^2  \biggl( {\partial \ln{\alpha} \over \partial x^i}
 - 2 {\partial \ln\phi\over\partial x^i} \biggr) = 0~~.
\label{hydromomcfa}
\end{eqnarray}
Here as in Eq.~(\ref{hydromom}), the
first term with ${\partial \alpha/\partial x^i}$ is
the relativistic analog of the  Newtonian gravitational force.

 In Eq.~(\ref{hydromomcfa}) there are two ways in which the
effective gravitational force might increase for finite $U^2$.
One is that the matter contribution to the source densities
for $\alpha$ or $\phi$
are increased by factors of 
$\sim 1 + U^2$ [cf. Eqs.~(\ref{phieq}) and (\ref{alphaeq})].
The more dominant effect, however,
 is from the  combination of the $S_j {\partial \beta^j /\partial x^i}$
term and the 
$U^2 [\partial \ln{\alpha} / \partial x^i - 
2\partial \ln{\phi}/\partial x^i]$
terms in Eq.~(\ref{hydromomcfa}).

As noted previously,   
these compression driving terms result from the affine connection part
$\Gamma^\mu_{\mu \lambda} T^{\mu \lambda}$ of the covariant differentiation
of $ T^{\mu \nu}$.  These terms have no Newtonian analog but
describe a general relativistic
increase in the gravitational force as 
$U^2$ increases.
As noted in \cite{wmm96,mw97} (see also Fig \ref{fig2} below)
for a binary, $U^2$ is approximately uniform over
the stars, and the increase in central density  due to these
additional forces scales as $\approx U^4$.  This scaling, however, 
is the net result from a nontrivial cancellation of terms and must be
treated carefully.  We shall return to this point below.

The proper way to determine the
post-Newtonian order at which the compression driving terms enter
 would be to count the powers of $c^2$ which appear in
the denominator of a term.  For example, if we divide the last two terms in
Eq.~(\ref{hydromomcfa}) by the gradient of the $\alpha$ term
(the analog of the Newtonian gravitational force) we would
obtain a ratio of order $U^2/c^2$ which would be
manifestly first post Newtonian.  However, in the
first post-Newtonian treatment of Wiseman \cite{wiseman}, these
velocity terms were explicitly disregarded.
Thus, the effects of these terms could not have been present
in that calculation.  It is no surprise, therefore that no effect
was observed in Ref. \cite{wiseman}.  

Also note that the 
$2{\partial \ln{\phi}/\partial x^i}$ term in Eq.~(\ref{hydromomcfa})
enters with
a sign such that the total $U^2$-dependent contribution is further increased 
by about twice that from the ${\partial \ln{\alpha}/\partial x^i}$
contribution alone.  (The factor of 2 in front of the derivative comes from 
the requirement that $\phi^2 \sim (1/\alpha)$ in the Newtonian limit 
\cite{wmm96}.)

A further increase of binding arises from the $K^{ij}K_{ij}$ terms in the
field sources, but these terms are much smaller than the
$U^2$ contributions for a binary system.

\subsection{Comment on the Relativistic Bernoulli Equation}

For comparison with other work in the literature it is
instructive to discuss the derivation of the relativistic 
Bernoulli equation from equation (\ref{hydromom}).  
It has been pointed out (e.g.~\cite{baumgarte}) 
that the hydrodynamics reduces to a simple equation 
for a fluid in which the velocity field  can be represented by 
a Killing vector. In our notation  this equation can be written,
\begin{equation}
d \ln{(U^t)} = {dP \over \sigma}~~.
\label{bernoulli}
\end{equation}

The demonstration that the relativistic Bernoulli equation 
(\ref{bernoulli}) is exactly reproduced from Eq.~(\ref{hydromom})
and indeed for any case in which a Killing vector 
can be imposed, was recently brought to 
our attention by T. Nakamura \cite{nakamura}.
We summarize the derivation  here in the conformally flat metric
both for clarity and to show that conformal flatness does
not violate this important constraint.

To begin with,
note that in the ADM formalism, the existence of
a Killing vector is equivalent to being able to choose
a shifted ADM grid such that $V^i = 0$
everywhere for the fluid.
Next use Eq.~(\ref{threevel})  to solve for $\beta^i$
and divide by $\sigma$.  The resulting
equation for stationary motion is
\begin{equation}
{1 \over \sigma} {\partial P\over \partial x^i} = U^t U_i {\partial \over \partial x^i}
\biggl({U_i \over \phi^4 U^t}\biggr)
 - (\alpha U^t)^2 {\partial ln{\alpha} \over \partial x^i}
 + 2 U^2  {\partial \ln{\phi} \over \partial x^i} ~~.
\label{bernoulli1}
\end{equation}
The recovery of the relativistic Bernoulli equation requires that
the r.h.s.~$= \partial \ln{U^t}/\partial x^i$.  With some straightforward
algebraic manipulation it is possible to show that  all of
the terms on the r.h.s.~cancel except for one term from the
$\beta$ derivative, $-\phi^{-4} U^2 \partial \ln{U^t}/\partial x^i$.
The completion of the proof is simply
to note that this term is equal to $\partial \ln{U^t}/\partial x$ by Eq.~(\ref{weq}).
The result is Eq.~(\ref{bernoulli}).
 
It is instructive to  consider the change in the
relativistic Bernoulli equation when there is no Killing vector,
i.e.~$V_i \ne 0$.  Along the same lines of the
derivation of Eq.~(\ref{bernoulli1}),
It can be shown \cite{nakamura}
that the momentum equation can be rewritten
in our notation as,
\begin{eqnarray}
{1 \over \alpha \sigma}\biggl[{\dot S_i}& + &  S_i{\dot \gamma \over\gamma}
+{1\over\gamma}{\partial\over\partial x^j}(S_iV^j\gamma)\biggr]
 + U^t U_j {\partial V^j \over \partial x^i} \nonumber \\
& =& 
-{1 \over \sigma} {\partial P\over \partial x^i}
+ {\partial ln U^t \over \partial x^i}~~.
\label{hydromomb}
\end{eqnarray}
The r.h.s.~is just the relativistic Bernoulli equation in
the limit that the l.h.s.~vanishes.  In general fluid flow,
however, the l.h.s.~contains not only the advection terms (in
brackets), but also an additional surviving part of
the $\beta^j$ derivative.  

It can be seen from this that 
imposing a Killing vector ($V^i = 0$) means 
that only simple hydrostatic
equilibrium is obtained for stationary systems.   However, when nontrivial hydrodynamic
motion is allowed, the extra forces embodied in the l.h.s.~of
Eq.~(\ref{hydromomb}) are manifest. 
The presence of fluid motion
not represented by a simple Killing vector, thus leads to
a deviation from the simple relativistic Bernoulli solution.
Any attempt to model this deviation requires 
a careful treatment of the dynamical properties of the
fluid described by the l.h.s.~of Eq.~(\ref{hydromomb}).

\section{CONSTRAINED HYDRODYNAMICS}
  
Further insight into the complexity of the physics contained 
in the relativistic equations of motion can be gained 
by considering some simple examples of constrained hydrodynamics
for which the answer is known.  These pose useful
tests of our numerical scheme.  Since some have proposed that the
effect we observe may be an artifact of numerical
resolution or approximation, we present here a summary
of various test problems designed to illustrate the
stability of the numerics and also to compare with
some of the calculations in the literature.  These
calculations also demonstrate that
the compression effect vanishes in the limiting
cases which have been studied by others.
Hence, they  could not have been observed.
They highlight the fact that the effect  we observe
only appears in a strong field dynamic 
treatment which accounts for internal  motion of 
stellar material in response
to the binary and its effect on the star's self gravity.
At present, ours may be the only existing result.
This is consistent with the conclusion of \cite{shapiro}
based upon test particle dynamics.

\subsection{Bench Mark Calculations}

To test for the presence of the compression driving
forces we consider two bench-mark initial
calculations.   
The bench mark of no compression 
is that of an isolated star.
In our three dimensional hydrodynamic calculations,
the single star structure is derived from Eq.
(\ref{hydromom}) in the limit 
\begin{equation}
S_i = U_i = V^i = \beta^i = 0~~.
\end{equation}
The condition of hydrostatic equilibrium in
isotropic coordinates is then trivially derived from
Eq.~(\ref{hydromom})
\begin{equation}
{\partial P\over \partial x^i} = - \sigma {\partial \ln{
\tilde \alpha} \over \partial x^i} ~~,
\label{static}
\end{equation}
where the tilde denotes that the metric coefficients are
evaluated in the fluid rest frame.
The Newtonian limit of the right hand side is recovered as 
$\tilde \alpha \rightarrow 1 - G m/r$.  Hence, we again
identify the ${\partial \ln{\tilde \alpha} / \partial x^i}$
term with the relativistic  analog of the Newtonian gravitational force.  
Eq.~(\ref{static}) also trivially reproduces to relativistic 
Bernoulli equation (\ref{bernoulli}).

We have of course tested our three dimensional calculations
for single isolated stars.  A single star remains stable on
the grid indefinitely, except when  baryon mass exceeds the
maximum stable mass allowed by the TOV equations. 
Above the maximum TOV mass the stars begin
to collapse on a dynamical timescale 
as they should.  We have also checked that 
the grid resolution used in our binary calculations 
is adequate to produce the correct central density,
stellar radius, and gravitational mass
of a single isolated star \cite{mw97}.  Hence, it seems unlikely that 
inadequate grid
resolution is the source of the compression effect as some have proposed.

In order to
facilitate comparisons with the literature, and to avoid 
confusion over equation of state (EOS) issues, 
we have employed a simplistic $\Gamma=2$ polytropic EOS,
$P = K\rho^\Gamma$, where $K= 1.8 \times 10^5$ erg cm$^3$ g$^{-2}$.
  This gives a maximum  neutron-star mass
of 1.82 $M_\odot$.
The gravitational mass of a single $m_B = 1.625$ $M_\odot$ star in isolation 
is 1.51 $M_\odot$ and the central density is 
$\rho_c =5.84\times 10^{14}$ g cm$^{-3}$.  The compaction ratio is  $m/R = 0.15$,
similar to one of the stars considered in \cite{baumgarte}.
Note that this EOS leads to stars with
a lower compaction ratio than
the stars we  considered in \cite{wmm96,mw97} 
for which $m/R \approx  0.2$.  
Hence, the effects of tidal forces in the present calculations 
should be more evident.

The bench mark in which compression
is present is that of two equal mass stars
in a binary computed with unconstrained hydrodynamics.
The binary  stars have the same baryon mass
($m_B = 1.625$ $M_\odot$ each), the same EOS, and
a fixed angular momentum $J = 2.5 \times 10^{11}$ cm$^2$
($J/M_B^2 = 1.09$ where M$_B = 2 m_B$).  For these conditions
the binary stars have $U^2 = 0.025$ and 
are at a coordinate separation of $\approx 100$ km.
The stars are stable but close to the collapse instability.  
Hence, they have 
experienced some compression which has increased their central
density by 14\% up to $\rho_c = 6.68\times 10^{14}$ g cm$^{-3}$.

The central densities of these two bench marks
are summarized in  the first and last entries of Table \ref{table1}. 
To compare with these bench-mark calculations
we have computed equilibrium configurations for stars under the
various conditions outlined below.
The test for the
presence or absence of compression inducing forces will
be the comparison of the numerically computed central density
with that of a single isolated star or stars in a binary.

\subsection{Stars in Uniform Translation}
\label{linear}

As a first nontrivial test, now
consider a star as seen from an observer in an 
inertial frame which is
in uniform translation  with respect to the fluid.
Choosing motion along the $x$ coordinate, the fluid three
velocity is,
\begin{equation}
 { U^x \over U^t} = V^x = {\rm Constant} ~~.
\end{equation}
However, the observer is still free to choose the ADM shift 
vector such that the computational grid  
remains centered on the star. 
 That is, although   $S_i, U_i, \ne 0$, we can still choose
$V^i = 0$. This gives a restriction on $\beta^i$ from Eq.~(\ref{threevel}),
\begin{equation}
\beta^x =   {\gamma^{x x}  \alpha  U_x \over W}  ~~. 
\label{betadef}
\end{equation}
Note, that this is an ADM coordinate freedom.  It is not
equivalent to a coordinate boost.  It is in fact a Killing vector
which is convenient for
numerical hydrodynamics.  It allows the matter to remain centered on the grid
even though the equations of motion are being solved for fluid 
which is not at rest with respect to the observer.

With $V^i = 0$, the  $x$ component of the momentum equation 
(in equilibrium) becomes,
\begin{equation}
{\partial P\over \partial x} = 
{S_x \over \alpha} {\partial \beta^x \over \partial x}
 - \sigma \biggl[(U^2 + 1)
 {\partial ln{\alpha} \over \partial x}
+ {U_j U_k  \over 2 } {\partial {\gamma^{j k}} \over \partial x} \biggr]~~.
\end{equation}
With a {\it CFC} metric this becomes,
\begin{equation}
{\partial P\over \partial x} = {S_x \over \alpha} {\partial \beta^x \over \partial x}
 - \sigma \biggl[(U^2 + 1)
 {\partial ln{\alpha} \over \partial x}
 - 2 U^2  {\partial \ln{\phi} \over \partial x} \biggr]~~.
\label{hydrostatcfa}
\end{equation}

There are now several differences between this expression
and that for an observer in the fluid rest frame.  
For one,  there is the shift vector
derivative $\partial \beta^x /\partial x$. Even in
uniform translation this derivative is nonzero
due to the variations of the metric coefficients over the star
(cf. Eq.~\ref{betadef}).
Also, the effective gravity is
enhanced by the $(U^2 + 1)$  velocity factor.
The ${U_j U_k  / 2 } {\partial {\gamma^{j k}} / \partial x^i}$ 
term also appears.  In addition, the effective source terms 
(\ref{phieq}) and (\ref{alphaeq}) for the
{\it CFC} metric coefficients are enhanced both by
 $(U^2 + 1)$ factors and the $K_{i j} K^{i j}$ term.

In spite of these differences, we nevertheless 
know that the locally determined
pressure and inertial density 
must be the same as those determined for a star at rest.
Indeed, since we can choose a Killing vector ($V^i = 0$)
 these equations must reduce to
the relativistic Bernoulli equation (\ref{bernoulli}).

Thus,  this is an
important numerical  test problem.   
We solve the full hydrodynamic equations explicitly,
e.g.~Eq.~(\ref{hydromomcfa}), under the initial condition of
nonzero $U_x$ for a single star. The cancellations 
embedded in the hydrodynamics are not
obvious.  Nevertheless, in the end, all of these effects must cancel to leave
the stellar central density unchanged (except for a Lorentz contraction
factor).

To solve the uniform translation problem numerically 
we have applied the Hamiltonian  and momentum constraints
 to determine the metric coefficients.
We then evolved the 
full hydrodynamic equations to equilibrium.
Figure \ref{fig2} shows the numerically evaluated 
central density for 
such translating stars as a function of $U^2$.
These stars were calculated with the EOS of \cite{wmm96}.
This is compared
with the central density for binary stars evolved at the
same $U^2$ value using the same EOS.  
One can see that the translating stars
maintain a constant central density (within numerical error)
as they should. In contrast, the central density
of binary stars grows as $\approx U^4$.
This growth is the nontrivial net result from the velocity dependent
terms in Eq.~(\ref{hydromomcfa}).   It is not obvious, however,
to what post-Newtonian order this dependence corresponds.

As summarized in Table \ref{table1},
the central density for a uniformly translating 
$\Gamma=2$ star with $U^2 = 0.025$ (for comparison with the 
binary  bench-mark calculation)
is $5.90 \times 10^{14}$ g cm$^{-3}$.  Within numerical
accuracy, this central density is identical 
with that of an isolated star at rest.

This is at least indicative that our 
observed growth in central density
may not be numerical error as some have suggested
(e.g.~\cite{brady,thorne97,shapiro}).
Such an error would likely be apparent in this test
case.  We argue that the difference between simple translation
and binary orbits relates to the physics of the binary 
system itself, in particular physics which is 
not apparent in uniform translation, an analysis of
tidal forces, 
or a truncated  expansion which does not contain sufficient
terms to adequately describe the
dynamical response of the fluid.

\subsection{Tidal Forces}
It has been pointed out \cite{lai,flanagan,thorne97}
that tidal forces are in the opposite sense to the compression
driving forces discussed here.  That is, tidal forces
distort the stars and decrease the central density and therefore
render the stars less susceptible to collapse.  We have
argued \cite{mw97} that although such stabilizing forces are present in 
our calculations they are much smaller in magnitude than
the velocity-dependent compression driving terms.  
Nevertheless, the evolution of the matter fields in
a calculation in which only tidal forces
are present still represents a useful  test of our 
numerical results.  Stars in which only tidal forces act,
should be stable
and the central density should
decrease rather than increase as the stars approach.

To test the effects of tidal forces alone
we have constructed an artificial test
calculation in which
 we place stars on the grid in a binary, but  with
no initial angular or linear momentum, 
i.e.~$J = 0$  and $U^2 = 0$.
This initial condition would normally
evolve to an axisymmetric collision between the stars.
However, after updating the matter fields,
we artificially return the center of mass of
the stars to the same fixed separation after each
time step.  We also reset to zero the mean velocity 
component directed along the
line between centers.  
This sequence is repeated until the matter fields 
come to equilibrium. Since the velocity
dependent forces eventually vanish, the
only remaining forces are the pressure and  static
gravitational (including tidal) forces.  

Results as a function
of separation distance are shown in Table \ref{table2} for the
$\Gamma=2$ EOS and Table
\ref{table3} for the realistic EOS used in \cite{wmm96}.
For the realistic EOS the central density indeed decreases 
as the stars approach consistent with the expectations
from Newtonian and relativistic tidal analyses \cite{lai,thorne97}.
For the $\Gamma=2$ polytropic EOS, the central density also
decreases as the stars approach and remain below the
central density of an isolated star.  The fact that this table
is not monotonic at the innermost point, however, 
may be due to a limitation
of this numerical approximation for tidal forces as the
ratio of separation to neutron-star radius diminishes.   

Although the tidal forces do indeed stabilize the stars, 
their effect on the central density
is quite small ($\sim 0.2\%$ decrease) compared to
the net increase in density caused by the compression forces
present for the binary.  This is consistent with the relative
order-of-magnitude estimates for these effects described in
\cite{mw97}.

\subsection{Stars in Rigid Corotation}
As a next nontrivial example, consider stars in a binary system which
are restricted to rigid corotation.  In a recent series of papers,
Baumgarte et al.~\cite{baumgarte} have studied neutron-star binaries
using the same conformally flat metric.  Their work differs from ours
in that rather than solving the hydrodynamic equations,
they describe the four velocity field by a Killing vector 
whereby the stars
are forced to corotate rigidly.  They also impose
spatial symmetry in the three Cartesian coordinate planes
so that they can solve the problem in only one octant.
One should keep in
mind, however, that 
rigid corotation is not necessarily the  lowest energy
configuration or the most natural 
 \cite{bc92} final state for two neutron stars near their 
final orbits.  This assumption, though  artificial, is 
nevertheless a means to
constrain and simplify the fluid motion degrees of freedom.
It is much easier to implement and therefore becomes an interesting
test problem for codes seeking to explore the true 
hydrodynamic evolution of close binaries.

Indeed, it is possible to show \cite{kramer80} that in
this limit, the neutron star hydrostatic equilibrium can be 
described by a simple Bernoulli equation in which
the compression driving   
force terms are absent except for a weak velocity dependence.  
Analytically, the reason for this is trivially obvious
from Eq.~(\ref{hydromomb}).  The existence of a Killing vector
is equivalent to setting $V^i = 0$ globally.  Choosing the
ADM coordinates to remain centered on the stars, 
in steady state the 
time derivatives vanish along with the rest of the l.h.s~of
Eq.~(\ref{hydromomb}).  Only the relativistic Bernoulli equation 
(\ref{bernoulli}) survives.

It is not surprising, therefore that in
 \cite{baumgarte} it has been demonstrated that in
this special symmetry, the central density of the stars does not
increase  (within numerical error) as the stars  approach the inner most
stable circular orbit.  In very close
orbits the density actually decreases relative to the central density
of stars at large separation.
They also find that the orbit frequency remains close to the
Newtonian frequency.
Both of these results are interesting in that they confirm 
that the compression effect does not occur (as it should not)
 in this special
symmetry.  They also demonstrate that conformal flatness
is not the source of the compression.

Accepting the results of \cite{baumgarte} as correct,  
this then becomes another important test of our calculations.
That is, if we artificially impose rigid corotation, then 
the central density should remain nearly constant until the
stars are close enough that tidal effects 
cause the central density to decrease rather than increase.

Imposing rigid corotation, however, is not a  trivial 
test problem to implement without completely replacing 
the hydrodynamic equations with the corresponding Bernoulli
solution of \cite{baumgarte,kramer80}.   (Indeed, we have done this
\cite{marronetti} and reproduce the results of \cite{baumgarte}
quite well).  Moreover,
we have found that directly modifying the 
hydrodynamic equations in an attempt to
mimic a dynamically unstable configuration
is difficult.   One might think
that the simplest way to implement corotation would be
to impose a high fluid viscosity.  Indeed high viscosity
would resist the hydrodynamic forces described herein.
However, a high fluid
viscosity also resists the much weaker tidal forces
and prevents the numerical relaxation to quasistatic equilibrium.  
It is thus difficult to achieve tidal locking by simply
increasing the viscosity.  

Instead, we introduce  artificial forces on the fluid
which continually drive the system toward  a state of
rigid corotation while allowing the system to at least 
somewhat respond hydrodynamically.  To do this
we define accelerations $(\dot U_i)_{\rm Rigid}$ 
necessary to achieve 
rigid rotation by
\begin{equation}
(\dot U_i)_{\rm Rigid} \equiv {(\tilde U_i - U_i) \over \Delta t}
\end{equation}
where $\tilde U_i$ are components of the rigidly corotating covariant four velocity
in the $ x- y$  orbit plane.  These are determined
by requiring that $\beta^i = (\omega \times  r)^i$
and setting $V^i = 0$ in Eq.~(\ref{threevel}).
\begin{equation}
\tilde U_y = {\omega x \phi^4 \over \alpha
\sqrt{1 - \omega^2  R^2 \phi^4/\alpha^2 }}~~,
\end{equation}
\begin{equation}
\tilde U_x = {-\omega y \phi^4 \over \alpha
\sqrt{1 - \omega^2   R^2 \phi^4/\alpha^2 }}~~,
\end{equation}
where $R$ is the coordinate distance from the center of mass of the binary.

At each time step we  then update the momentum density
using a combination of the hydrodynamic and corotating
acceleration terms,
\begin{equation}
\dot U_i = f (\dot U_i)_{\rm Rigid} + (1 - f)(\dot U_i)_{\rm Hydro}
\end{equation}
where $(\dot U_i)_{\rm Hydro}$ is the acceleration from
the full hydrodynamic equation of motion [Eq.~(\ref{hydromomcfa})].

Numerically, we find that 
if $f$ is small ($< 0.2$)  the hydrodynamic forces
dominate and corotation is not obtained.
On the other hand, for $f > 0.2$ the system is not stable, 
i.e.~the stars deform and the velocities become erratic. 
 We have therefore  run with
$f = 0.2$ which temporarily 
produces a velocity field which is close to rigid corotation.  
That is, the residual three velocities 
are damped to a fraction of the orbit 
speed.  This is, perhaps, good enough to make qualitative comparisons
with the expectations from a truly corotating 
system.  

Starting from the unconstrained initial
configuration, we find that when the stars
have achieved approximate corotation
the central density has decreased from 
$6.68 \times 10^{14}$ g cm$^{-3}$ to
$5.90 \times 10^{14}$ g cm$^{-3}$ which is
close to the value for stars in isolation 
($5.84 \times 10^{14}$ g cm$^{-3}$).
The calculated gravitational mass is slightly greater
than that of the unconstrained binary.  
However, with the large artificial
force terms needed to approximate corotation, gravitational
mass is not a well defined quantity in this simulation.
Also, the orbit frequency was not 
 sufficiently converged for a meaningful comparison.

\subsection{The Spin of Binary Stars}
As noted above our simulations indicate that 
neutron stars relax to a state of almost no intrinsic
spin.
In a separate paper \cite{wm98} we analyze the nature
and formation of this state in more detail.  
For the present discussion, however, we summarize
in Figure \ref{fig3} 
a study of the relaxation to this state from states
of arbitrary initial rigid rotation (including corotation).

As a means to distinguish the intrinsic spin motion of the fluid
with respect to a non-orbiting distant observer, we 
define a quantity which is analogous
to volume averaged intrinsic stellar spin in the orbit plane,
\begin{equation}
J_S = \sum_{i=1,2}\int \biggl[(x - \tilde x_i)S_y - (y-\tilde y_i)S_x \biggr] 
{\phi^2 \over \alpha} d V_i~~,
\end{equation}
where $(\tilde x_i,\tilde y_i,\tilde z_i = 0)$ is the coordinate
center of mass of each  star.  

In this study we have imposed an initial  angular velocity 
$\omega_S$ in the corotating frame 
 to obtain various initial rigidly rotating spin angular momenta
 (including corotation, $\omega_S = 0$), but for fixed  total
$J/M_B^2 = 1.4$.
We have considered spin angular frequencies in the
range  $-900 < \omega_S < 900$ rad sec$^{-1}$, corresponding
to $-0.03 < J_S/m_B^2 < 0.17$.  We then let the system
evolve hydrodynamically with the stars maintained at zero temperature.

 In Ref.~\cite{mw97} we showed that neutrino emission
is sufficient to radiate away the released gravitational
energy and keep the stars at near zero temperature until just before collapse.
This is the reason that we have treated this as a relaxation
problem.  That is, unlike a true hydrodynamic calculation, the 
relaxation calculation presented here,  assumes
that the stars radiate efficiently and stay at zero temperature.  
Therefore, this evolution does
not need to conserve energy or circulation.  
This relaxation assumption is the reason
the stars can evolve to a different spin (lower energy)
state without violating the circulation theorem.

Figure \ref{fig3} shows the spin  $J_S/m_B^2$ as a function of time for
each initial condition.  In each case, the system relaxed to
a state of almost no net spin within about
three sound crossing times ($t \sim 0.6$ msec).
These calculations suggest that rapidly spinning 
neutron stars in close orbits are unstable.  The true evolution 
time, however, would be much longer.

We also note that the quantity $\int \sigma [\sqrt{1+ U^2}
-1] dV$
decreased as the system evolved from
rigid rotation to hydrodynamic equilibrium.
Since this quantity is related to the kinetic energy of the binary, 
this indicates that the hydrodynamic lowest energy state
is one of lower kinetic energy (for fixed total
angular momentum) than that of rigid rotation.

As far as the compression effect is concerned, 
one wishes to know whether the response of the stars
is simply due to that fact that they have no
spin (and therefore no internal centrifugal 
force to support them
against the compression forces), or whether more complex
fluid motion within the star itself affects the stability.
To test this, we have constructed stars of no
spin ($J_S = 0$) by simply damping the residual
motion to that of $J_{S} = 0$ after each 
update of the velocity 
fields.  

Since this  no-spin state
is so close to the true hydrodynamic equilibrium, 
this produced stable $J_{S} = 0$ equilibrium stars for the binary.
For this case, the central density converges
to $\rho_c = 6.56 \times 10^{14}$ g cm$^{-3}$ which
is very close to the high value for the unconstrained
hydrodynamics.  This result would seem to indicate that
most of the increase in density can be attributed to
the velocity with respect to the corotating frame
generated by the fact that the stars have almost no spin.

\section{DISCUSSION}

For clarity, we summarize in this section our conclusions
regarding why the neutron-star compression  effect was not observed
in some other recent works. 

First consider post-Newtonian expansions.
In the work of Wiseman \cite{wiseman} the force terms containing
$U^2$ were explicitly deleted from the computation of the stellar 
structure [cf.~Eq.~(8) in that paper].  
Only the $dln{\alpha}/dx$ term was included.
The recovery of simple hydrostatic equilibrium was thus unavoidable.

The PN orbiting ellipsoids of Shibata  et al.~\cite{shibata} included
more terms.  Indeed, it was noted that
 there are two effects at 1PN order.  One is the 
self gravity of each star of the binary and the other
is the gravity acting between the stars.  In their calculations
the self gravity dominates causing the stars to become more compact.
This is  consistent
with the compression effect described here in the sense that relativistic
corrections can dominate over Newtonian tidal forces.
However, the self gravity terms in \cite{shibata} appear to only include
the usual 1PN terms which would equally apply to stars in isolation.
Hence, the velocity-dependent compression driving terms are probably not  present.

Their results for
stars in corotation are consistent with ours under the same constraint.
They also note that approaches in which PN corrections to
the gravity between the stars is included without also including
the corrections to the self gravity (as in \cite{lw96}) can be misleading.

In the work of Lombardi et al.~\cite{lombardi} both
corotating  and irrotational equilibria
were computed.  However, in their calculations it appears
that the stars become less compact as they approach contrary to 
our results and the results
of \cite{shibata}.  It may be that the reason for this is
that in Lombardi et al.~the post-Newtonian corrections to
self gravity were only computed for stars "instantaneously at rest".
The authors chose to "exclude the spin kinetic energy contribution
to the self energy".  It is such terms, however, 
which we identify with the compression effect.

The conformally flat corotating equilibria
computed by Baumgarte et al.~\cite{baumgarte} are consistent
with our results.  Since their stars were restricted to rigid corotation,
only the hydrostatic Bernoulli solution would result.  They could
not have observed the compression forces which result from fluid motion
with respect to the corotating frame.

We have argued in this paper that if one wishes
to explore this effect,  it would be best to apply
a complete unconstrained  strong-field relativistic hydrodynamic
treatment for stars which are not in corotation. 
In this regard, a recent paper \cite{sbs98}  has come to our attention
in which hydrodynamic simulations of 
both corotating and irrotational binaries have been studied
in a first post-newtonian approximation to conformally-flat
 gravity but using the full relativistic hydrodynamics
equation (\ref{hydromom}).  For both  corotational and irrotational
stars the central
density is observed to oscillate about a value which is less than
that of isolated stars. Hence, the authors conclude that no compression
effect is present.  

Since this calculation contains  many of the higher order terms to which we attribute the
compression effect, it is
not immediately obvious why the compression effect was not observed.
This may indeed be a real contradiction.  We suggest, however,  that
this simulation did not observe the effect because of their use of an
unrealistically soft $\Gamma = 1.4$ EOS.   The authors chose this EOS because
the stars become so extended that one can compute arbitrarily close binaries 
without encountering  the relativistic inner orbit instability. 
For the irrotational stars (model $Bc$ in \cite{sbs98}), which is the only
simulation that might have observed
the compression effect, the compaction ratio is only $M/R = 0.023$.  Hence,
a 1.45 M$_\odot$ neutron star would have an unrealistic  radius of 93 km.      

However, since they have simulated very extended stars at very
close separation,  the tidal forces are
much stronger relative to the relativistic compression driving terms than in any of
the simulations which we have done.  

The ratio of the stabilizing tidal correction $\Delta E_{tidal}$
to the destabilizing energy from compression $\Delta E_{comp}$ should scale 
\cite{mw97,lai} as  
\begin{equation}
{\Delta E_{tidal} \over \Delta E_{comp}} \propto  \biggl({R \over r}\biggr)^6~~,
\end{equation}
where $R$ is the neutron star radius and $r$ is the orbital separation.
For model Bc in \cite{sbs98} we estimate that  this ratio is $^>_\sim 200$
times greater than any of the binary stars we have considered.  Hence, it is quite likely
that the authors have simply chosen an unrealisticly soft equation of state for which
the tidal forces dominate over compression.  It might be very interesting to
see the results from a similar study for stars with a realistic compaction ratio
and several radii apart.

Concerning tidal expansions, in Brady \& Hughes \cite{brady} an 
attempt was made to analyze
the stability of a central star perturbed by an orbiting point
particle.  The metric and stress-energy were perturbed in terms
of order $\epsilon = \mu/R$ where $\mu$ is the point particle mass and $R$ its 
coordinate distance from the central star.  The Einstein equation was
then linearized to terms of order $\epsilon$.  The result of
this linearization was that the only possible correction
to the central density was a  single  monopole term of order $\mu/R \sim v^2$.
However, in our numerical results as shown in Figure \ref{fig3}.  
the central density is observed to increase as $v^4$.
Hence, it may be that the expansion of Ref. \cite{brady} was truncated
at too low order to observe the compression effect described here.
The main reason that they could not observe the effect, however,
is that the terms involving motion of the central star were discarded.
We attribute the compression effect to an enhancement of the self
gravity due to motion of the stars with respect to the
 corotating frame.  Hence,
the neglect of terms involving motion of the central star precludes the
possibility of observing the effect.

We believe that the same conclusion is true in the treatments by 
Refs. \cite{flanagan,thorne97}.
The analysis of Flanagan
\cite{flanagan}  is based upon
the method of matched asymptotic expansion.
The metric is approximated
\begin{equation}
g_{\mu \nu} = \eta_{\mu \nu} + h_{\mu \nu}^{NS} +  h_{\mu \nu}^B~~,
\end{equation}
where the superscript $NS$ refers to the self contribution from one star and $B$
refers to the contribution from a distant companion.
The internal gravity of a static neutron star  $h_{\mu \nu}^{NS}$
is expanded to all orders.
The binary  tidal contribution $h_{\mu \nu}^B$ 
is expanded  in powers of
the ratio of stellar radius to orbital separation.

First we suggest that such a decomposition 
may be questionable for 
a close neutron-star binary.  In our metric one can write the
metric perturbation as 
\begin{equation}
h_{i j} = (\phi^4 -1) \delta_{i j}~~.
\end{equation}  
The conformal factor $\phi$ is a solution to a Poisson equation
involving source terms from the two stars.  Between the
stars, the only source of the fields arises
from the $K_{i j} K^{i j}$ terms which are quite small.
Hence, neglecting $K_{i j} K^{i j}$ terms,
 $\phi$ is additive in the "vacuum" between the stars,
\begin{equation}
\phi = \phi_1 + \phi_2  =   1 +  
{m_1 \over 2 \vert r-r_1 \vert} 
 +   {m_2 \over 2 \vert r-r_2 \vert} ~~.
\end{equation}
Expanding $h_{i j}$  around star $1$ in the
 presence of a distant companion $2$ we have
\begin{eqnarray}
h_{i j}& =& {4 \over 2}\biggl({m_1 \over \vert r-r_1 \vert} + 
{m_2 \over \vert r-r_2 \vert}\biggr)\nonumber \\
& & +  {6 \over 4}\biggl({m_1 \over \vert r-r_1 \vert} + 
{m_2 \over \vert r-r_2 \vert}\biggr)^2 + \cdot \cdot \cdot \nonumber \\
&&\nonumber \\
&& =  h_{\mu \nu}^{NS} +  h_{\mu \nu}^B + {\rm cross~terms} ~~.
\end{eqnarray}
However, for the binary systems we have considered,
the cross terms are $\sim$ 15\% to 20\% of the sum 
$h_{\mu \nu}^{NS} + h_{\mu \nu}^B$.  Hence, they can not
be neglected.  
The errors associated with this decomposition may be 
part of the reason that
the compression effects are not apparent in this work.

A related concern is with the expansion of the stress-energy
tensor in \cite{flanagan}. 
We have noted 
that most of the compression arises from the net effect of
velocity dependent terms in the covariant derivative of the stress-energy
tensor.  
In \cite{flanagan} the stress energy is expanded is powers
of the curvature $R^{-m}$.  
The author  states  
\cite{flanagan} "We assume initial conditions
of vanishing $T_{\mu \nu}^{(2)}$, so that the only source
for perturbations is the external tidal field."  
An  analysis which only considers perturbations from
the external tidal field (and not motions of the fluid) 
will not observe the compression effect.
The result  of \cite{flanagan} is that the central
density is unchanged until tidal forces enter at $O(R^6)$.
This is consistent with our results in the limit of only tidal perturbations 
acting on the stars.  It is not clearto us, however, to what degree the
velocity dependent terms are included or excluded by this
expansion.  A more careful recent revision (E. Flanagan, Priv. Comm.) 
shows an effect coming in a lower order, but not necessarily
as strong as we have noted.

In the paper of Thorne \cite{thorne97}, a similar tidal expansion 
is applied.  In that work only the stabilizing effect
of tidal forces was considered  along with the stabilizing effect
of rotation. However, the increased self gravity
from velocity-dependent forces was not included.  Hence, the 
conclusions of \cite{thorne97} 
are consistent with our results based upon tidal forces.
So are the Newtonian tidal effects computed in \cite{lai}.

\section{CONCLUSIONS}

The results of this study (cf.~Table \ref{table1}) are that we
see almost no difference 
between the central density of an isolated star and a binary star
in which rigid corotation has been artificially 
 imposed, or one in which
only tidal effects are included.  Indeed, in the case
of tidal forces alone, the central density in our simulations
 actually decreases as stars approach, consistent with
other works.

An increase in the central density
is only apparent in our binary simulations for stars
with fluid motion with respect to the corotating frame.
(Specifically we considered stars of low  intrinsic spin in a binary.)  
In such cases
there is no simple Killing vector which can be imposed
to cancel the compression driving forces.
We have argued here and in \cite{mw97} that the 
main compression effect arises from the net result of velocity-dependent
hydrodynamic terms \cite{fn1}.  These terms arise from
the affine connection part of the covariant differentiation
of the stress-energy tensor. 

We show here that the compression 
effect would not have been observed in a study of tidal
forces or any model which artificially imposes rigid corotation
of the fluid.
A proper treatment must consider
all of the force terms apparent in the momentum equation (\ref{hydromom})
to sufficient order that their effects on the fluid self gravity survive.
A similar conclusion has been reached in \cite{shapiro}
based on test particle dynamics near a Schwarzschild black hole.
In that work it is concluded that at least 2.5 post-Newtonian particle
dynamics is necessary before a dynamical  collapse instability is
manifest.  

We argue that the results of this study are thus
consistent with  results in a number of recent papers
\cite{lai,rs96,wiseman,shibata,lombardi,lw96,brady,flanagan,thorne97,baumgarte} which have analyzed the stability
of binary stars in various approximations and limits and see no effect.   
Since we do not disagree with the lack of a compression effect
in the limits which they
have imposed, we conclude that the existence or
absence of the neutron-star
compression effect has not yet been independently tested.

Therefore, 
if one wishes
to explore this effect,  it would be best to apply
a complete unconstrained  strong-field relativistic hydrodynamic
treatment employing an EOS which produces realistically compact
neutron stars. 
Another alternative, however, might 
be to study the quasi-equilibrium structure
of nonspinning irrotational  binary  stars at sufficiently high order.
In this regard a  recently proposed formalism \cite{bonazzola}  to compute 
quasi-equilibria for nonsynchronous binaries may be
of some use.  We have begun calculations in this independent formalism.
The results will be reported in a forthcoming paper.

Regarding the existence of this low spin state, 
we find that such a state
represents the unconstrained hydrodynamic equilibrium for a close binary.  
In Newtonian theory, stars are driven to corotation by
tidal forces.  However in \cite{bc92} it has been shown
that Newtonian tidal forces are insufficient to produce corotation
before neutron-star merger unless the
viscosity is unrealistically high.  Nevertheless,
in the absence of 
strong tidal forces, neutron stars stars gradually spin down. 
Therefore, even apart from
the hydrodynamic effects described here, stars of low
spin are likely to be members of close binaries.
The hydrodynamic effects described herein, however, 
could hasten the spin down as stars approach their final orbits
and cause the stars to become more compact.

\acknowledgments

Work at University of Notre Dame
supported in part by DOE Nuclear Theory grant DE-FG02-95ER40934,
NSF grant PHY-97-22086, and
by NASA CGRO grant NAG5-3818.
Work performed in part under the auspices 
of the U.~S.~Department of Energy
by the Lawrence Livermore National Laboratory under contract
W-7405-ENG-48 and NSF grant PHY-9401636.

\begin{figure}
\caption{The scaled Cotton-York tensor component $C^{\theta \phi} m^3$
as a function of proper radius $r/m$ for a maximally rotating $a = m$ 
Kerr black hole.  This quantity is a measure
of the deviation from conformal flatness.}
\label{fig1}
\end{figure}

\begin{figure}
\caption{Numerically evaluated central density for
a uniformly translating star (lower curve) as a function of $U^2 \equiv
 U^i U_i$.  This is compared
with the central density for binary stars (upper curve) with the
same average $U^2$ value. These calculations utilized the EOS of
Ref. [2].}
\label{fig2}
\end{figure}

\begin{figure}
\caption{Intrinsic neutron star spin $J_S/m_B^2$
as a function of coordinate time.
The curves are labeled by the  initial angular velocity
$\omega_S$ (in units of 100 rad sec$^{-1}$) relative to the corotating frame.
}
\label{fig3}
\end{figure}

\vfill\eject

\begin{table}
\caption{Central density for m$_B = 1.625$ M$_\odot$ stars in various
conditions using a $\Gamma = 2$ EOS. }
 \begin{tabular}{ccc}
 Environment & Constraints &
$\rho_c$ ($10^{14}$ g cm$^{-3})$ \\
&&\\
 \tableline
&&\\
Single Star & Hydrostatic & $5.84$ \\
Single Star & Uniform Translation& $5.90$ \\
Binary & Tidal Only & $5.82$ \\
Binary & Rigid Corotation & $5.90$ \\
Binary & Rigid No Spin & $6.56$ \\
Binary & Full Hydrodynamics & $6.68$ \\
\label{table1}
\end{tabular}
\end{table}
 
\begin{table}
\caption{Central density vs.~coordinate separation between centers for
m$_B = 1.625$ M$_\odot$  ($\Gamma=2$)  stars in which only tidal forces are included. 
The neutron star radius (in isotropic coordinates) is 12 km.}
 \begin{tabular}{ccc}
 Separation  (km) & 
$\rho_c$ ($10^{14}$ g cm$^{-3})$ \\
&&\\
 \tableline
&&\\
41.8   & $5.821$ \\
50.6   & $5.816$ \\
81.8   & $5.821$ \\
$\infty$  & $5.837$ \\
\label{table2}
\end{tabular}
\end{table}

\begin{table}
\caption{Same as Table II but for the EOS of ref.~[2].
The neutron star radius (in isotropic coordinates) is 6 km.}
 \begin{tabular}{ccc}
 Separation  (km) & 
$\rho_c$ ($10^{14}$ g cm$^{-3})$ \\
&&\\
 \tableline
&&\\
31.2   & $14.15$ \\
37.4   & $14.16$ \\
64.8   & $14.20$ \\
103.8  & $14.25$ \\
$\infty$  & $14.30$ \\
\label{table3}
\end{tabular}
\end{table}


\begin{references}


\bibitem{wm95}J.R. Wilson and G.J. Mathews, Phys. Rev. Lett., {\bf 75}, 4161 (1995).

\bibitem{wmm96}J.R. Wilson, G.J. Mathews, \& P. Marronetti, Phys. Rev. D{\bf 54}, 1317 (1996).

\bibitem{mw97}G. J. Mathews and J. R. Wilson, Astrophys. J, {\bf 482}, 929 (1997).

\bibitem{lai}D. Lai, Phys. Rev. Lett., {\bf 76}, 4878 (1996).

\bibitem{rs96}R. Reith and G. Sch\"afer,  Phys. Rev. D. submitted (1996).

\bibitem{wiseman}A. G. Wiseman,  Phys. Rev. Lett., {\bf 79}, 1189 (1997).

\bibitem{shibata}M. Shibata, Prog. Theo. Phys., {\bf 96}, 317 (1996); Phys. Rev.
D{\bf 55}, 6019 (1997); K.  Taniguchi \& M. Shibata, Phys. Rev.
D{\bf 56}, 798 (1997); M. Shibata \& K.  Taniguchi, Phys. Rev. D{\bf 56}, 811 (1997).;
M. Shibata,  K.  Taniguchi \& T. Nakamura, Prog. Theo. Phys., (1998) submitted.

\bibitem{lombardi}J. C. Lombardi, F. A. Rasio, \& S. Shapiro, Phys. Rev. 
D{\bf 56}, 3416 (1997).

\bibitem{lw96}K. Taniguchi \& T. Nakamura, Prog. Theor. Phys., {\bf 96}, 
693 (1996);
D. Lai \& A. G. Wiseman, Phys. Rev., D{\bf 54}, 3958 (1996); W. Ogawaguchi \& 
Y. Kojima, Prog. Theor. Phys., {\bf 96}, 901 (1996).

\bibitem{brady}P. Brady and S. Hughes, Phys. Rev. Lett., {\bf 79}, 1186
(1997).

\bibitem{flanagan}E. Flanagan, Phys. Rev. D., submitted (1997) (gr-qc)/9706045).

\bibitem{thorne97}K. Thorne, Phys. Rev. D., submitted (1997)  gr-qc/9706057.

\bibitem{baumgarte}T. W. Baumgarte, G. B. Cook, M. A. Scheel, 
S. L. Shapiro \& S. A. Teukolsky, Phys. Rev. Lett., {\bf 79}, 1182  
(1997) gr-qc/9704024; also gr-qc/9705023; gr-qc/9709026.

\bibitem{sbs98}M. Shibata,, T. W. Baumgarte \& S. L. Shapiro,  Phys. Rev. D, Submitted
(1998) gr-qc/9805026.

\bibitem{w79}J. R. Wilson, in {\it Sources of Gravitational 
Radiation}, ed. L . Smarr (Cambridge; Cambridge Univ. Press) p. 423 (1979).

\bibitem{adm62}R. Arnowitt, S. Deser, and C. W. Misner, in {\it Gravitation},
ed. L. Witten (New York: Wiley), p. 227 (1962).

\bibitem{york79}J. W. York, Jr., in {\it Sources of Gravitational 
Radiation}, ed. L . Smarr (Cambridge; Cambridge Univ. Press) p. 83 (1979).

\bibitem{evans85}C. R. Evans, PhD. Thesis, Univ. Texas (1985).

\bibitem{cook93}G. B. Cook, M. W. Choptuik, M. R. Dubal,
S. Klasky, R. A. Matzner and S. R. Oliveira, Phys. Rev. D{\bf 47}, 1471 (1993).


\bibitem{brugmann97}B. Br\"ugmann, Phys. Rev. Lett., {\bf 78}, 3606 (1997).

\bibitem{smarr}P. Anninos, D. Hobill, E. Seidel, and L. Smarr,
Phys. Rev. Lett., {\bf 71}, 2851 (1993).

\bibitem{abrahams}A. M. Abrahams, in {\it
Sixth Marcel Grossmann Meeting, Kyoto 1991}, H. Sato, T. Nakamura, eds.,
World Scientific: Singapore) p. 345 (1992).

\bibitem{mtw}C. W. Misner, K. S. Thorne, and J. A. Wheeler, {\it Gravitation},
(San Francisco: W. H. Freeman and Co.) (1973).
 
\bibitem{thorne}K. S. Thorne, Rev. Mod. Phys., {\bf 52}, 299 (1980).

\bibitem{kramer80}D. Kramer, H. Stephani, E. Herlt, M. MacCallum,  in
{\it Exact Solutions of Einstein's Field Equations},
ed. E. Schmutzer (Cambridge University Press: Cambridge), (1980).

\bibitem{eisenhart}L. P. Eisenhart,
{\it Riemannian Geometry},
(Princeton University Press: Princeton), (1966).

\bibitem{cookcfa}G. B. Cook, S. L. Shapiro, S. A. Teukolsky,
Phys. Rev.  D{\bf 53}, 5533 (1996).

\bibitem{york73}J. W. York, Jr., J. Math. Phys., {\bf 14}, 456.

\bibitem{nakamura} T. Nakamura 1997, Priv. Comm.

\bibitem{shapiro}S. L. Shapiro, Phys. Rev. D{\bf 57}, 1084 (1998).

\bibitem{marronetti} P. Marronetti, G. J. Mathews  and J. R. Wilson, 
Phys. Rev. Lett., {\it Submitted} (1998), gr-qc/9803093.

\bibitem{bc92} L. Bildsten \& C. Cutler Ap. J., {\bf 400}, 175 (1992).
 
\bibitem{wm98}J.R. Wilson and G.J. Mathews, Phys. Rev. Lett., {\it
submitted} (1997).

\bibitem{fn1} Although in 
\cite{mw97} we pointed out that velocity dependent corrections 
enter into the source for the post-Newtonian potential this was
for illustration and not identification 
of the main effect.

\bibitem{bonazzola}S. Bonazzola, E. Gourgoulhon, and J.-A. Marck,
Phys. Rev. D{\bf 56}, 7740 (1997); S. A> Teukolsky, Ap. J. Submitted,
gr-qc/03082 (1998); M. Shibata, Phys. Rev. D, Submitted, gr-qc/03085 (1998);
E. Gourgoulhon, Phys. Rev. D, Submitted, gr-qc/04054 (1998).


\end{references}
\end{document}